\documentclass[12pt]{book}

\usepackage[dvips]{graphicx,color}
\usepackage{makeidx,physics,cosmology}

\makeauthorindex
\makeindex

\BookTitle{Frontier in Astroparticle Physics and Cosmology}
\CopyRight{\copyright 2004 by Universal Academy Press, Inc.}

\begin{document}

\BookTitle{\itshape Frontier in Astroparticle Physics and Cosmology}
\CopyRight{\copyright 2004 by Universal Academy Press, Inc.}
%\tableofcontents
\pagenumbering{arabic}

\chapter{%   %%%%%%%%% <===== TITLE of the contribution
%%%%%%%%%%% The first letter of each word should be capital letter.
Cosmological Origin of Small-Scale Clumps \\
and DM Annihilation Signal}

\author{%
     Veniamin BEREZINSKY \\%%%%%% <== First author
{\it  Laboratori Nazionali del Gran Sasso, Istituto Nazionale di Fisica
 Nucleare \\
I-67010  Assergi (AQ), Italy}\\
     Vyacheslav DOKUCHAEV  \\%%%%%% <== Second author
{\it Institute for Nuclear Research of the Russian Academy of Sciences, \\
60th Anniversary of October Prospect 7a, Moscow 117312, Russia} \\
     Yury EROSHENKO \\%%%%%% <== Second author
{\it Institute for Nuclear Research of the Russian Academy of Sciences,\\
60th Anniversary of October Prospect 7a, Moscow 117312, Russia} }

\AuthorContents{V.\ Berezinsky, V.\ Dokuchaev and Yu.\ Eroshenko}
\AuthorIndex{Berezinsky}{V.}
\AuthorIndex{Dokuchaev}{V.}
\AuthorIndex{Eroshenko}{Yu.}

\section*{Abstract}

We study the cosmological origin of small-scale DM clumps in the
hierarchical scenario with the most conservative assumption of adiabatic
Gaussian fluctuations.  The mass spectrum of small-scale clumps with $M
\leq 10^3 M_{\odot}$ is calculated with tidal destruction of the clumps
taken into account within the hierarchical model of clump structure.
Only $0.1-0.5$\% of small clumps survive the stage of tidal destruction
in each logarithmic mass interval $\Delta\ln M\sim1$.  The mass
distribution of clumps has a cutoff at $M_{\rm min}$ due to diffusion of
DM particles out of a fluctuation and free streaming at later stage.
$M_{\rm min}$ is a model dependent quantity.  In the case the
neutralino DM, considered as a pure bino,  $M_{\rm min}
\sim 10^{-8} M_{\odot}$.  The evolution of density profile in a DM clump
does not result in the singularity because of formation of the core
under influence of tidal interaction.  The radius of the core is $R_c
\sim 0.1 R$, where $R$ is radius of the clump.  The applications for
annihilation of DM particles in the Galactic halo are studied.  The
number density of clumps as a function of their mass, radius and
distance to the Galactic center is presented.  The enhancement of
annihilation signal due to clumpiness, valid for arbitrary DM particles,
is calculated. In spite of small survival probability, the global
annihilation signal in most cases is dominated by clumps, with major
contribution given by small clumps. The enhancement due to large
clumps with $M\geq 10^6 M_{\odot}$ is very small.

\section{Introduction}

The gravitationally bound structures in the universe are developed from
primordial density fluctuations $\delta(\vec x,t)=\delta\rho/\rho$.  They
are produced at inflation from quantum fluctuations.  The predicted power
spectrum of these fluctuations has a nearly universal form $P(k)\propto
k^{n_p}$, with $n_p\simeq1$.  At radiation-dominated epoch the
fluctuations grow logarithmically slowly.  After transition at $t=t_{\rm
eq}$ to the matter-dominated epoch, the fluctuations grow as
$\delta\propto t^{2/3}$.  The gravitationally bound objects are formed
and detached from cosmological expansion when fluctuations enter the
non-linear stage $\delta\geq1$.  The non-linear stage of fluctuation
growth has been studied both by analytic calculations \cite{Zeldovich} and
\cite{ufn1} and in numerical simulations \cite{NFW,moore99,JingSuto} for
Large Scale Structure (LSS).  The density profile in the inner part of
these objects is given by $\rho(r) \propto r^{-\beta}$, with with $\beta
\approx 1.7 - 1.9$ in analytic calculations \cite{ufn1}, $\beta =1$ in
simulations of NFW \cite{NFW} and $\beta =1.5$ in simulations of Moore et
al.  \cite{moore99} and Jing and Suto \cite{JingSuto}.  In this work we
apply this approach to the smallest DM objects in the universe, which we
shall call {\em clumps}.  The clumps, being the smallest structures, are
produced first in the universe, and it makes difference of our
consideration from LSS formation.  The theoretical observation of this
work is the importance of tidal interaction in the process of DM clump
formation:  the central nonsingular core is formed in the clumps and large
fraction of clumps are tidally disrupted.

We use in the calculations the hierarchical model, in which due to
merging of objects a small clump is hosted by the bigger one, the latter
is submerged to more bigger etc.  We use the standard cosmology with
WMAP parameters.  The primordial spectrum index is $n_p=0.99 \pm 0.04$
(WMAP) or $n_p=0.93 \pm 0.03$ (WMAP+2dF+Ly$\alpha$).

\section{Tidal Destruction of  Clumps in Hierarchical Model}
\label{destruction}

The destruction of clumps by the tidal interaction occurs at the epoch of
their hierarchical formation, long before the formation of galaxies.
This interaction arises when two clumps pass near each other and when a
clump moves in the external gravitational field of the bigger host to
which this clump belongs.  In both cases a clump is exited by the external
gravitational field, i.~e.  its constituent particles obtain additional
velocities in the c.~m.  system.  The clump is destroyed if its internal
energy increase $\Delta E$ exceeds the corresponding total energy $|E|
\sim GM^2/2R$.  In \cite{bde03} we have calculated the rate of excitation
energy production by both aforementioned processes.  The dominating
process is given by tidal interaction in the gravitational field of the
host clumps, with the main contribution from the smallest host clump.  We
use the Press-Schechter formalism \cite{ps74,cole} for hierarchical
clustering.
A small-scale clump during its life can be a constituent part of many
host clumps of successively larger masses. After tidal disruption
of the lightest host, a small clump becomes a constituent part of the
larger host etc. Transition of a small clump from one host to another
occurs nearly continuously in time up to the formation of a big enough
host, where tidal destruction becomes inefficient.

The fraction of mass in the form of clumps which escape the tidal 
destruction in each logarithmic mass interval $\Delta\ln M\sim1$ is 
found as
\begin{equation}
\xi_{\rm int}\simeq0.01(n+3).
\label{xitot}
\end{equation}
In other words the mass function of clumps is $\xi_{\rm int} dM/M$.
Since $n$ is close to $- 3$, only a small fraction of clumps about
$0.1-0.5$\% survive the stage of tidal destruction.  However, this
fraction is enough to dominate the total annihilation rate in the
Galactic halo.

\section{Core of Dark Matter Clump}
\label{core}

We use the following parameterization of the density profile in a
clump:
\begin{equation}
 \rho_{\rm int}(r)=\left\{ \begin{array}{lr}
 \rho_c, & r<R_c; \\ \displaystyle{
 \rho_c\left(\frac{r}{R_c}\right)^{-\beta} },
 & R_c<r<R;\\ 0, & r>R.
 \label{rho}
 \end{array} \right.
\end{equation}
In \cite{ufn1} the relative core radius of the clump is estimated
as $x_c=R_c/R\simeq\delta^3_{\rm{eq}}\ll1$ from consideration of
the perturbation of the velocity field due to damped mode of the
cosmological density perturbations.  Here $\delta_{\rm{eq}}$ is
an initial density fluctuation value at the end of radiation
dominated epoch.  In \cite{BBM97} the core is considered to be
produced for spherically symmetric clump by inverse flow caused by
annihilation of DM particles.  We show \cite{bde03} that these
phenomena are not the main effects and that much stronger
disturbance of the velocity field in the central part of clumps is
produced by tidal forces.  The tidal forces influence the nearly radial
motion of  DM particles at the time of clump formation.  As a result
these particles obtain some angular momentum which prevents the
formation of singularity.
Once the core is produced it is not destroyed in the evolution followed.
The core formation proceeds mainly near the time of the clump maximal
expansion $t_s$. At this moment the clump decouples from an expansion
of the universe and contracts in the non-linear regime.  Soon after this
period a clump enters the hierarchical stage of evolution, when the tidal
forces can destroy it, but surviving clumps retain their cores.

The calculations proceed in the following way (see \cite{bde03} for
details).  The background gravitational field (including
that of the host clumps) is expanded in series in respect to the
distance from the point with maximum density in a fluctuation.  The
motion of a DM particle in this field is studied.  The spherically
symmetric term of the expansion causes the radial motion of a particle
in the oscillation regime.  Spherically non-symmetric term describes the
tidal interaction.  It results in deflection of a particle trajectory
from a center (point with maximum density).  The average (over
statistical ensemble) deflection gives the radius of the core $R_c$.
After statistical averaging, $R_c$ is expressed through the amplitude of
the fluctuation $\delta_{\rm eq}$ and the variance $\sigma_{\rm eq}$ (or
$\nu=\delta_{\rm eq}/\sigma_{\rm eq}$) as
\begin{equation}
 x_c=R_c/R\approx 0.3 \nu^{-2}f^2(\delta_{\rm eq}),
 \label{x_c}
\end{equation}
where the function $f(\delta_{\rm{eq}})\sim1$ is given in Ref \cite{bde03}.
The fluctuations with $\nu \sim 0.5 - 0.6$ have $x_c \sim 1$, i.e.  they
are practically destroyed by tidal interactions.  Most of galactic clumps
are formed from $\nu \sim 1$ peaks, but the main contribution to the
annihilation signal is given by the clumps with $\nu\simeq2.5$ for which
$x_c\simeq0.05$.

\section{Clumps of Minimal Mass}
\label{smmin}

The mass spectrum of clumps has a low-mass cutoff at $M=M_{\rm min}$,
which value is determined by a leakage of DM particles from the overdense
fluctuations in the early universe.  CDM particles at high temperature
$T>T_f \sim 0.05 m_{\chi}$ are in the thermodynamical (chemical)
equilibrium with cosmic plasma.  After freezing at $t>t_f$ and $T<T_f$,
the DM particles remain for some time in {\em kinetic} equilibrium with
plasma, when the temperature of CDM particles $T_{\chi}$ is equal to
temperature of plasma $T$.  At this stage the CDM particles are not
perfectly coupled to the cosmic plasma.  Collisions between a CDM particle
and fast particles of ambient plasma result in exchange of momenta and a
CDM particle diffuses in the space.  Due to diffusion the DM particles
leak from the small-scale fluctuations and thus their distribution obtain
a cutoff at the minimal mass $M_D$.

The DM particles get out of the kinetic equilibrium when the energy
relaxation time for DM particles $\tau_{\rm rel}$ becomes larger than the
Hubble time $H^{-1}(t)$.  This conditions determines the time of kinetic
decoupling $t_d$.  At $t \geq t_d$ the CDM matter particles are moving in
the free streaming regime and all fluctuations on the scale of
free-streaming length $\lambda_{fs}$ and smaller are washed away.  The
corresponding minimal mass $M_{\rm fs} =
(4\pi/3)\rho_{\chi}(t_0)\lambda_{\rm fs}^3$, is much larger than $M_D$ and
therefore $M_{\rm min}=M_{\rm fs}$.  In \cite{bde03} we have performed the
calculations using two methods:  the transparent physical method, based on
the description of diffusion and free streaming, and more formal method
based on solution of kinetic equation for DM particles starting from the
period of chemical equilibrium.  Both methods agree perfectly.  The
minimal mass in the DM mass distribution is determined by the process of
free-streaming.  For the case of neutralino (bino) as DM particle this
minimal mass equals to $M_{\rm min} = 1.5\times10^{-8}M_{\odot}$ for
neutralino mass $m_{\chi}=100$~GeV and the mass of selectron and sneutrino
$\tilde{M}=1$~TeV.  Our calculations agree reasonably well with that of
\cite{bino01}, while $M_{min}$ from \cite{gzv12} coincides with our value
for $M_D$.

\section{Annihilation Signal Due to  Small Clumps}
\label{enhancement}

There is distribution of clumps in the Galactic halo over three
parameters, mass $M$, radius $R$, and distance from the Galactic Center
$l$:  $n_{\rm cl}(M,R,l)$.  This distribution also depends on the
parameters which describe the internal structure of the clumps, $\beta$
and $x_c=x_c(M,R)$, from Eq.~(\ref{rho}).  With the number
density of clumps  in the halo written as $dN_{\rm{cl}} =
n_{\rm{cl}}(l,M,R)d^3ldMdR$,  the observed signal at the position of the
Earth from DM particle annihilation in the clumps is given by
quantity
\begin{eqnarray}
I_{\rm cl}&=&\frac{1}{4\pi}
\int\limits_{0}^{\pi}d\zeta\sin\zeta
\int\limits_{0}^{r_{\rm{max}}(\zeta)}\frac{2\pi r^2dr}{r^2}
\int\limits_{M_{\rm{min}}}^{M_{\rm{max}}}dM
\int\limits_{R_{\rm{min}}}^{R_{\rm{max}}} dR \nonumber \\
&& \times\,\, n_{\rm{cl}}(l(\zeta,r),M,R) \dot N_{\rm cl}(M,R),
\label{ihal}
\end{eqnarray}
where $r$ is distance from the Sun (Earth) to a clump and $\zeta$ is angle
between the line of observation and the direction to the Galactic center,
$r_{\rm{max}}$ is the distance from the Sun to the halo's outer border
and $\dot N_{\rm cl}(M,R)$ is annihilation rate in the single clump of
mass $M$ and radius $R$.

Additional annihilation signal is given by unclumpy DM in the halo
with homogeneous ({\em i.e.} smoothly spread) density $\rho_{\rm
DM}(l)$, where $l$ is a distance to the Galactic Center.
\begin{equation}
I_{\rm{hom}}=\frac{\langle\sigma_{\rm ann}v\rangle}{2}
\int\limits_{0}^{\pi}\!
d\zeta\sin\zeta\!\!\!\int\limits_{0}^{r_{\rm{max}}(\zeta)}
\!\!\!\!dr\rho_{\rm{DM}}^2(l(\zeta,r))/m_{\chi}^2. 
\label{hom}
\end{equation}
The {\em enhancement} $\eta$ of the signal due to a presence of
clumps is given by
\begin{equation}
\eta=\frac{I_{\rm cl}+I_{\rm{hom}}}{I_{\rm{hom}}} \label{eta}
\end{equation}
This quantity describes the global enhancement of the
annihilation signal observed at the Earth ({\em e.~g.} the flux
of radio, gamma, and neutrino radiations) as compared with usual
calculations from annihilation of unclumpy DM.
\begin{figure}[t]
 \begin{center}
 \includegraphics[height=21.5pc]{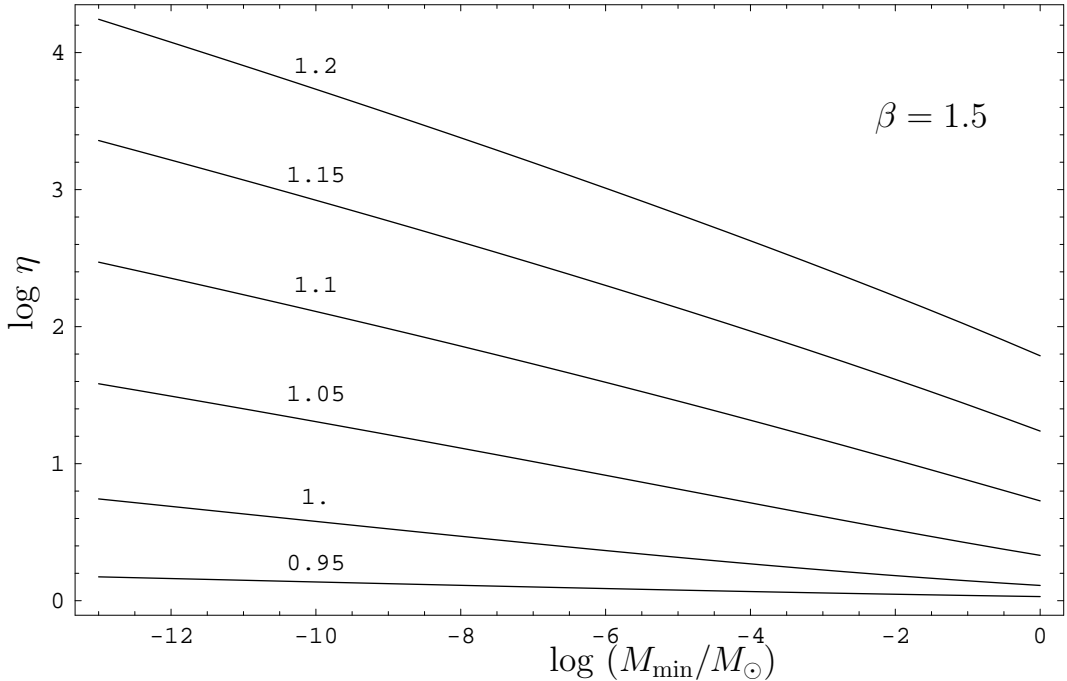}%% ``13pc'' is just the example.
  \end{center}
 \caption{The global enhancement $\eta$ of the annihilation signal
from the Galactic halo as a function of the minimal clump mass
$M_{\rm{min}}$, for clump density profile with index $\beta=1.5$  and for
different indices $n_p$ of primeval perturbation spectrum.  The curves are
marked by the values of $n_p$.}
 \label{fig3r}
\end{figure}
We assume that space density of clumps in the halo, $n_{\rm cl}(l)$ is
proportional to the unclumpy DM density, $\rho_{\rm DM}(l)$:
$n_{\rm{cl}}(l)=\xi\rho_{\rm DM}(l)/M$ with $\xi \ll 1$.  This assumption
holds with a good accuracy for the small-scale clumps.
The signal from small clumps is determined
mainly by clumps of the minimal mass.  In calculations \cite{bde03}
we used different
density profiles in the clumps, the distribution of DM clumps over their
masses $M$ and radii $R$, and the distribution of clumps in the galactic
halo.  The enhancement depends on the nature of DM particle only through
$M_{\rm min}$. For $\beta=1.5$ enhancement is given in
Fig.~\ref{fig3r} Details of calculations and the plots for other
values of $\beta$ can be found
in \cite{bde03}.  The enhancements $\eta$ for $n_p=1$ or less is
not large:  typically it is 2-5 for $M_{\rm min}
\sim 10^{-8}M_{\odot}$.  For example, $\eta=5$ for $n_p=1.0$ and
$M_{\rm{min}}= 2\cdot 10^{-8}M_{\odot}$.  It strongly increases for smaller
$M_{\rm{min}}$ and larger $n_p$.  For example, for $n_p= 1.1$ and
$M_{\rm{min}}=2\cdot 10^{-8}M_{\odot}$, enhancement is
$\eta=130$.  Our approach is based on
the hierarchical clustering model in which smaller mass objects are formed
earlier than the larger ones, i.~e.  $\sigma_{\rm{eq}}(M)$ diminishes with
the growing of $M$.  This condition is satisfied for objects with mass
$M>M_{min} \simeq 2\cdot10^{-8}M_{\odot}$ only if the primordial power
spectrum has the power index $n_p>0.84$.  The
enhancement of the annihilation signal is absent, e.g  $\eta\simeq1$, for
$n_p<0.9$.

\section{Annihilation Signal Due to  Big Clumps}
\label{enhancement1}

Numerical simulations reveal in the galactic halos the big clumps with
masses $10^8-10^{10}M_{\odot}$. At $ t \sim t_{\rm eq}$ these clumps 
are characterized  
by an effective power spectrum $P(k) \propto k^{n}$ with
$n\approx -2$  (in contrast to $n\approx -3$
for small clumps), and thus the survival probability given by
Eq.~(\ref{xitot}) is larger for the big clumps. Indeed, the effective power
spectrum index
\begin{equation}
n= -3\left[1+2 \,\frac{\partial\ln \sigma_{\rm eq}(M)}{\partial \ln
M}\right]
 \label{n}
\end{equation}
tends to $n= n_p-4=-3$ for small $M$ and to $n \approx -2$ for
$M$ in the interval $10^6 - 10^9 M_{\odot}$. On the other hand
the mean internal number density of DM particles in the big clumps is much
smaller in comparison with that in the small clumps, and it compensates
the first effect.

The number density $n(M)$ of the big clumps with $M\geq 10^8 M_{\odot}$
in the numerical simulations found \cite{moore99} as
$n(M)dM \propto dM/M^{\gamma}$ with $\gamma \approx 1.9 - 2.0$ and with
a mass fraction of clumps $\varepsilon \sim 0.1 - 0.2$. Observations of
halo lensing \cite{clobs} give smaller values $\varepsilon \sim 0.06-0.07$.
It is interesting to note that the mass function of clumps, obtained
from Eq.~(\ref{xitot}), is  close (including the normalization coefficient)
 to that
obtained in the numerical simulations for big clumps with mass
$M\geq10^8M_{\odot}$. Strictly speaking our calculations are not valid for
big clumps, because of their destruction in the halo up to the present
epoch $t_0$ and accretion of new clumps into the halo. Nevertheless, for
the small interval of masses, where the power-law spectrum can be used as a
rather good approximation, our approach appears to be roughly valid.

Calculations of the enhancement factor $\eta$ from simplified 
Eqs.~(\ref{ihal}) and (\ref{hom}) are
performed by using $\varepsilon= 0.1$, the number density distribution of
clumps in the Galaxy $n_{\rm cl}(l) \propto \rho_{\rm DM}(l)$, and internal
density distribution of the DM particles in clumps
\begin{equation} \rho(r)=\rho_c (r/a)^{-\beta}(1+r/a)^{-\kappa}
\label{profile}
\end{equation}
valid down to the core radius $R_c$. The core is defined as $\rho(r)
 =\rho_c=const$
at $r \leq R_c$. The NFW profile has $\beta =1$ and $\kappa = 2$, while the
Moore et al. profile has $\beta=\kappa=1.5$.

The other important parameter is clump radius $R$, which determines
the average density $\bar{\rho}$ at the epoch of DM virialization in the
clump. This quantity is calculated from the value of overdensity $\delta$
at the linear stage of clump formation. The values of $\delta$ have
Gaussian distribution and the normalization of fluctuation spectrum was
performed as usual to the value of r.m.s. fluctuation at the 8 Mpc scale
$\sigma_8\simeq1$. This approach corresponds to the picture that the 
big clumps in the Galactic
halo are similar to the small protogalaxies (galactic building
blocks), which escape the tidal destruction when capturing by the
Galaxy. The tidal stripping of the outer parts of the clumps change their
structure. Nevertheless this process is not important for clumps with
mass $M\ll 10^{10} M_{\odot}$. For minimal and maximal masses of the 
big clump we use $M_{\rm min}=10^8 M_{\odot}$ and 
$M_{\rm max}=10^{10} M_{\odot}$.  

As a result of
calculations, identical to those  in Section 5, we  have found the  
enhancement factor $\eta=1.02$ for the NFW profile \cite{NFW}  
with $\gamma=1.9$, $\varepsilon=0.1$, $R_c=a$, $R/a=5$ (the case of 
existence of the core), and  $\eta=1.14$ for the case of absence of
the core $R_c=0$. 

For the Moore et al.
profile \cite{moore99},  $\eta=1.06$ for 
$\varepsilon=0.1$, $R_c/a=0.5$, $R/a=5$ and 
$\eta=1.16$ for a smaller core $R_c/a=0.2$.  

Diminishing of $M_{\rm min}$ increases the enhancement weakly,
approximately as $\eta(M_{\rm min}) \propto M_{\rm min}^{-0.35}$.

We conclude that enhancement of the annihilation signal due to the big clumps
is small.

We are grateful to  Ben Moore for providing us with the data which allow us
to normalize the density distribution of DM in the clumps.

\end{document}